\documentclass{article}
\usepackage{amsmath,amssymb,mathrsfs,amsfonts,dsfont}
\usepackage{pdflscape}
\usepackage{afterpage}
\usepackage[utf8]{inputenc}
\usepackage{graphicx}
\usepackage[numbers]{natbib}
\usepackage{tikz}
\usetikzlibrary{positioning}
\usepackage[T3,OT1]{fontenc}
\DeclareSymbolFont{tipa}{T3}{cmr}{m}{n}
\DeclareMathAccent{\invbreve}{\mathalpha}{tipa}{16}

\newlength{\defbaselineskip}
\newcommand{\setlinespacing}[1]%
           {\setlength{\baselineskip}{#1 \defbaselineskip}}

\newcommand{\singlespacing}{\setlength{\baselineskip}{\defbaselineskip}}

\begin{document}
\begin{center} {\LARGE\textbf{Narratives of Quantum Theory \\ \vskip .3em in the Age of Quantum Technologies}}
\vskip 2em
{\large \bf Alexei Grinbaum} \\
{\it CEA-Saclay/IRFU/LARSIM, 91191 Gif-sur-Yvette, France
\par Email alexei.grinbaum@cea.fr}
\vskip 1em 
\end{center}


\abstract{\noindent Quantum technologies can be presented to the public with or without introducing a strange trait of quantum theory responsible for their non-classical efficiency. Traditionally the message was centered on the superposition principle, while entanglement and properties such as contextuality have been gaining ground recently. A less theoretical approach is focused on simple protocols that enable technological applications. It results in a pragmatic narrative built with the help of the resource paradigm and principle-based reconstructions. I discuss the advantages and weaknesses of these methods. To illustrate the importance of new metaphors beyond the Schr\"odinger cat, I briefly describe a non-mathematical narrative about entanglement that conveys an idea of some of its unusual properties. If quantum technologists are to succeed in building trust in their work, they ought to provoke an aesthetic perception in the public commensurable with the mathematical beauty of quantum theory experienced by the physicist. The power of the narrative method lies in its capacity to do so.
}

\section{Introduction}

Historically, there are two ways to present quantum theory to the public. The first begins with experimental results that cannot be explained classically. Such experiments typically involve very small or very large spatial scales. When quantum theory is called to explain them, it is perceived as a story about ``two infinities,'' accounting for phenomena at the microscale (e.g., the two-slit experiment) or the macroscale (e.g., gravitational waves). The term ``two infinities'' is much older than contemporary physics: it goes back to Pascal's \textit{Pens\'ees} and the Greeks but never fails to impress even in our time. Traditional introductions of the second kind begin with a theoretical rather than an experimental fact, most frequently the superposition of quantum states. This trait is usually presented with the help of the Schr\"odinger cat metaphor. Since mathematical formalism cannot be used in a popular account, comprehension is not sought in either kind of introduction. Instead one pursues the goal of getting the audience accustomed to the strangeness of quantum theory. The public should learn to \textit{live with} it~\cite{dupuygrinbaumnano}. A similar goal, according to Freeman Dyson, is also pursued in the introductory courses of quantum mechanics despite their use of mathematical methods:
\begin{quote}
The student begins by learning the tricks of the trade.\ldots This is the first stage in learning quantum mechanics, and it is comparatively painless. The second stage comes when the student begins to worry because he does not understand what he has been doing. He worries because he has no clear physical picture in his head. He gets confused in trying to arrive at a physical explanation for each of the mathematical tricks he has been taught.\ldots This second stage is strenuous and unpleasant. Then, unexpectedly, the third stage begins. The student suddenly says to himself, ``I understand quantum mechanics,'' or rather he says, ``I understand now that there isn’t anything to be understood.'' The difficulties which seemed so formidable have mysteriously vanished. What has happened is that he has learned to think directly and unconsciously in quantum-mechanical language.\ldots The duration and severity of the second stage are decreasing as the years go by. Each new generation of students learns quantum mechanics more easily than their teachers learned it.\ldots There is less resistance to be broken down before they feel at home with quantum ideas. Ultimately, the second stage will disappear entirely. Quantum mechanics will be accepted by students from the beginning as a simple and natural way of thinking, because we shall all have grown used to it. By that time, if science progresses as we hope, we shall be ready for the next big jump into the unknown.~\cite{Dyson58}
\end{quote}
When Dyson wrote this in 1958, he was apparently convinced the problem would simply get dissolved itself over time. The difficulty, however, has all but disappeared. Instead, many alternative views of quantum mechanics emerged that brought a challenge for the very understanding anyone may have had of quantum mechanics in 1958. They have also deeply modified the ``quantum-mechanical language.'' 

A different but equally theoretical approach to presenting quantum mechanics appears in the years following Dyson's article. It is based on entanglement rather than superposition and is decidedly more modern since it follows the developments in the foundations of quantum theory after 1964, when the ``second quantum revolution''~\cite{AspectBell} occurred with the publication of John Bell's inequalities~\cite{Bell1}. However, merely replacing one enigmatic property: the superposition of the states of quantum systems, with another equally enigmatic one: entanglement between the states of two subsystems of a composite quantum system, cannot help to reduce the aura of mystery around quantum mechanics. A decisive insight comes from the reconstruction program~\cite{grinbjps}.

Since von Neumann's and Birkhoff's work on quantum logic~\cite{bvn} axiomatic reconstructions have been our primary way to single out the key feature or features of quantum theory. An axiomatic reconstruction is a derivation of the formalism from a set of principles with a clear meaning. Contrary to the old-fashioned interpretations of quantum mechanics, axiomatic reconstructions possess supplementary persuasive power provided by mathematical derivation. Formal results are established as valid theorems; one cannot suspect them to be \textit{ad hoc}. `Why is it so?'---`Because we derived it.' The question of meaning, previously asked with regard to the formalism, now bears on the selection of first principles. Already in the 1970s some favoured this approach as a new foundation for teaching quantum theory. For example, Land\'e wishes for a ``systematic explanatory approach based on knowledge  of the `principal reason behind the quanta' (Einstein)''~\cite{Lande}. The importance of explanation became particularly obvious in the reconstructions put forward in the last two decades. After Hardy's seminal work~\cite{hardy} it became \textit{oblig\'e} to explain the significance of each postulate via a catchword or a brief phrase in natural language: ``Probabilities,'' ``Simplicity,'' ``Composition,'' or ``Continuity.'' This clarity of meaning comes in stark contrast with previous decades when an abstract formulation of the axioms would have been sufficient. Notable old examples include Mackey's axioms~\cite{mackey57} or Ludwig's highly complicated approach~\cite{ludwig}. For years mathematical formulation used to prevail over physical meaning; now meaning takes center-stage.

Two lessons of the reconstruction program: the use of derivation and its focus on meaning, are the two central components of a majority of contemporary narratives about quantum technologies. The latter often start with a simple protocol or a resource. In one or several steps, the protocol is shown to lead to the realization of a device or a resource to be spent on doing useful work. This is a pragmatic rather than a theoretical approach. It is constructed along the lines of the resource paradigm instead of being centered on a fundamental concept. Its main message is about performing work or making a calculation rather than expounding on the strangeness of a non-classical property. This is supposed to make it easier for the public to learn to \textit{live with} quantum technologies.

The importance of the pragmatic approach for the pedagogy of quantum mechanics can hardly be overestimated, yet it would be an exaggeration to say that the potential of this method for presenting quantum theory to the public has been sufficiently exploited. I will review some of its applications and discuss its advantages and weaknesses (Section~\ref{sect_wonder}). I will then turn back to the theoretical approach and single out entangelement (Section~\ref{sect_entangle}). If entanglement and the associated notion of the amount of non-locality are key properties that need to be communicated to the public, then we need to replace the Schr\"odinger cat by a new metaphor to enable a new narrative of quantum theory. I will suggest a set of comparisons for this problem in the context of two theoretical accounts of the composition of parts in the history of philosophy. In Section~\ref{sect_myth}, I will discuss the problem of transmitting to the public a sense of beauty that practising physicists perceive directly when they employ the mathematical formalism of quantum theory. This is the \textit{hard problem} of popularization (by analogy with the hard problem of consciousness~\cite{Chalmers,chalm}). I submit that, if the hard problem is not solved, one should not expect trust in quantum technologies from society at large. Trust is not exclusively a matter of rational argument or pragmatic efficiency. It depends on the affective link to be established between the public and the scientist as a person. While this link obviously relies on the rational content of a narrative, it is primarily mediated through the symbolic and aesthetic dimensions. I will propose a narrative of entanglement with a potential to promote trust, which uses myth as another medium that combines strangeness and beauty. Although the aesthetics of myth is not mathematical, a carefully constructed analogy may succeed both in conveying the pragmatic interest of quantum technologies and in eliciting in the lay audience a sense of what quantum physicists experience when they plunge into the formalism of quantum theory.

\section{Wonder and pragmatism}
\label{sect_wonder}

It is instructive to list various features promoted by popular science writers or elementary textbook authors as being a central non-classical property of quantum physics. The role of major source of wonder about quantum mechanics has been traditionally played by the Schr\"odinger cat metaphor~\cite{schro35}, implying that one such feature was the superposition of quantum states. Dirac did not use metaphor but still emphasized the importance of superposition~\cite{dirac}; Feynman also built his more popular lectures around the superposition principle. The uncertainty principle and the notion of complementarity were used by the ``founding fathers,'' respectively Werner Heisenberg and Niels Bohr but also Wolfgang Pauli and others, as two other key non-classical ideas demarcating the point of departure between classical and quantum physics. Standard textbooks written between the 1930s and the beginning of 1970s by Fock~\cite{fock} or Landau~\cite{LL} in the Soviet Union, Messiah~\cite{Messiah} or Cohen-Tannoudji~\cite{CTtextbook} in France, and in English by Tomonaga~\cite{Tomonaga} employ the correspondence principle and the classical-quantum relation to facilitate the transition from classical to quantum concepts.
Wave-particle duality is often found at the center of such approaches, although emphasizing it has also been criticized at the time, e.g, by Tomonaga. In a remarkable historic review Rechenberg names these two properties as two foundational stones of the most popular approach to teaching quantum mechanics~\cite{Rechenberg}. He then proceeds with describing two other methodologies based, respectively, on the notion of probability and on path integrals for quantum electrodynamics. R\"udinger leans towards Rechenberg's second option, namely the foundational methods based on axioms. He argues that despite being demanding on students in terms of time and creative thinking, the approach that begins with the origin of the Hilbert space and probability is the only psychologically fulfilling and profound teaching method~\cite{Rudinger}. Such was the essence of the 1970s debate on quantum didactics.

Six years before Rechenberg wrote his 1970 review, John Bell formulated a crucial non-classical feature of quantum physics in a different way. This new approach would take three or four decades to penetrate into textbooks and teaching. Bell's inequality put forward the tension exemplified in the concept of quantum entanglement between classical notions of realism and locality. The term `entanglement' (Verschr\"ankung) had been introduced by Erwin Schr\"odinger thirty years earlier~\cite{schro35bis,schro36} but Bell was the first to use it in a quantitative criterion of non-classicality~\cite{Bell1}.

Contemporary approaches part ways with the choices made and with the metaphors employed by the founding fathers of quantum theory. This is obviously not because their methods and concepts had become wrong but because, after so many years, they are seen as old-fashioned and incapable of capturing or transmitting the essence of the new work done in the disciplines of quantum foundations and quantum information. As a result, emphasis has shifted away from complementarity or superposition; these notions, although they do not totally disappear from popular accounts or elementary textbooks, are typically relegated to second place. Instead the authors put forward one of the new thrilling features that refer to more recent breakthroughs in understanding quantum theory (Table~\ref{table_authors}). One can grossly divide them into several groups:
\begin{description}
  \item[Structural features] Compositional structure in category-theoretic models~\cite{coecke_struct}, convex operational models~\cite{barrett}, and general informational foils~\cite{Foils}. Tensor product structure of quantum theory~\cite{barnum_generalized_2007}. The amount of non-locality in quantum theory and postquantum models~\cite{van_dam_nonlocality_2000,linden_quantum_2007,pawlowski_information_2009,Masanes}. Measures of non-classicality other than quantum entanglement, i.e., discord, steering, or contextuality.
  \item[Computational protocols] Quantum teleportation~\cite{teleport93}, no cloning~\cite{WoottersCloning}, quantum key distribution~\cite{BB84,Eckert1991}, Shor's algorithm~\cite{Shor}, and other non-classical computational and/or cryptographic features. No-go theorems expressed as impossibility of certain protocols or tasks~\cite{Bub,bubstudies}. Device-independent approaches~\cite{MayersYao,grin_devindep}.
  \item[New common-language paradoxes] Timeless formalisms and postselection paradoxes, weak measurements, contradictory traits such as negative occupation numbers~\cite{HardyParadox1,HardyParadox2,AharonovHardy}, `Cheshire cat' properties~\cite{AharonovCheshire}, or unintuitive particle trajectories~\cite{threeslits}.
\end{description}

\begin{table}
  \centering
  \begin{tabular}{|ll|}
  \hline
  \textbf{Non-classical feature} & \textbf{Example of author} \\ \hline
  Uncertainty & W. Heisenberg \\
  Complementarity & N. Bohr \\
  Quantum superposition & E. Schr\"odinger \\
  Entanglement and non-locality & J. Bell \\
  \hline
  Tensor product structure & B. Coecke and A. Kissinger~\cite{PicturingBook}\\
  Amount of non-locality & J. Bub~\cite{BubBanana} \\
  Cloning and teleportation & A. Zeilinger~\cite{ZeilingerPhotons} \\
  Quantum discord & V. Vedral~\cite{VedralBook} \\
  Quantum contextuality & D. Mermin~\cite{MerminBook}, A. Cabello~\cite{CabelloSing} \\
  Quantum randomness & N. Gisin~\cite{GisinBook} \\
Cryptographic protocols & V. Scarani~\cite{ScaraniBook2,ScaraniBook} \\
  Time and postselection paradoxes& Y. Aharonov and D. Rohrlich~\cite{ARbook}
\\  \hline
\end{tabular}
  \caption{Notion used by various authors to convey a sense of strangeness about quantum theory. The first group shows the notions used by the ``founding fathers,'' while the second group illustrates some modern approaches.
The selection from recent popular and semi-popular literature about quantum theory is not intended to be complete and is provided solely for illustration.}\label{table_authors}
\end{table}

It is of particular interest that popular and semi-popular accounts focusing on technologies, i.e., on quantum computing and quantum cryptography, rarely or never venture to explain the entire road from a background theoretical notion (superposition, entanglement, discord, steering, or contextuality) to actual technology. Such an endeavor seems perhaps too complex. A shortcut is to begin directly with non-intuitive but simply formulated quantum protocols, i.e., no cloning of quantum information, teleportation of quantum states, or various no-go theorems in quantum computing. Such elementary protocols then serve as building blocks in the introduction of more complicated quantum protocols, devices, or technology. This method is a spin-off of the reconstruction program of quantum theory~\cite{grinbjps}. Unlike traditional interpretations that were heaped over the existing quantum formalism, reconstructions seek to derive the formalism from simple physical principles or axioms:
\begin{quote}
Quantum mechanics will cease to look puzzling only when we will be
able to \textit{derive} the formalism of the theory from a set of
simple physical assertions (``postulates,'' ``principles'') about
the world. Therefore, we should not try to append a reasonable
interpretation to the quantum mechanical formalism, but rather to
\textit{derive} the formalism from a set of experimentally
motivated postulates.~\cite{RovRQM}
\end{quote}
Reconstructions combine arguments of simplicity and clarity with a mathematical derivation. They offer a new view of the structure of the theory: an element of its formalism, e.g., the Hilbert space, is not introduced axiomatically but emerges as a consequence of certain assumptions, e.g., postulates about the amount of relevant information or how information can be used~\cite{BZ,grinbijqi,Chiri}. This view opens an invaluable perspective for education and popularization of quantum theory. Instead of struggling with approximate and often metaphoric introductions of mathematical terms, authors can now focus on the logical link between a simple principle and an element derived from it. If the derivation can be at least schematically presented, the message for the lay audience or for undergraduate students becomes very clear: it is the meaning of the principle which is responsible for an otherwise obscure mathematical construction. The introductions to quantum technologies based on elementary protocols build on this pedagogical insight from the reconstruction program.

Protocols demonstrate what can be done; they follow a pragmatic approach even if the implication is that one has to renounce to seek deeper understanding. The results of such protocols, while remaining conceptually mysterious, are perfectly tangible and can be implemented empirically. A telling example of the approach focused on tangible protocols is a pictorial version of quantum theory proposed by Coecke and Kissinger~\cite{coecke_pict,PicturingBook}. On this view, the complete quantum theory is replaced by a set of structural elements necessary for the introduction of computational protocols. For teleportation, for example, only matters the tensor product structure of the composition of subsystems of a composite system, while the underlying single-system descriptions are relegated to second place. The authors now remove the underlying mathematics of quantum theory and propose a general category-theoretic framework exclusively devoted to the properties of composition. It transpires that a set of protocols can be introduced in this model without further assumptions needed for obtaining full-blown quantum theory. This mathematically unusual but pragmatically justified \textit{tour de force} is arguably more than an operational approach to quantum theory, because the latter need not appear at the end of the development. When used in a semi-popular account (e.g.,~\cite{CoeckeKindergarten}), the pictorial view fully replaces quantum theory with a hands-on, what-can-be-achieved toolkit. That quantum protocols or technologies can be so introduced, without any need for presenting quantum theory proper, is a \textit{novum} among popular and semi-popular accounts. In previous work, it was necessary to explain theoretic notions in order to capture the essence of counterintuitive experimental findings. In the newer accounts, by contrast, the strangeness of fundamental quantum theory is entirely relegated to, or even replaced by, one or several non-classical protocols.

The upshot of such accounts is a remarkable departure from the traditional ways of popular science. To get a sense of quantum technology, it is no more necessary to question the meaning of quantum theory nor to venture into its fundamental tenets. The methodological argument runs as follows: since quantum mechanics is difficult or impossible to be understood by laypersons, who have no mastery of its mathematical formalism, it is best to altogether drop the goal of achieving such understanding. Instead, an introduction to quantum technology may start from a set of pragmatic communication protocols. On the scientific side, these protocols, e.g., quantum teleportation, are theorems of quantum theory; a textbook account would be normally grounded in the knowledge of underlying science. However, in a popular account simple quantum protocols often become elementary building blocks that convey to the audience a pragmatic, hands-on heuristic while abandoning all intent to present the mathematical formalism of quantum theory.

An immediate advantage of the methodology based on protocols is that authors are able to limit their demonstration to one deductive step, either logical or technological. The public is presented with something simple. For example, monogamy of entanglement is shown to lead in just one big step to the realization of a quantum key distribution device; or discord becomes the enabler of quantum cryptography. The audience then perceives these features as resources provided by quantum theory. However strange they may appear, resources can be used whenever present. The workings of the technological device appear as a legitimate and logical consequence of spending such resources. This approach is in line with several contemporary views on quantum theory and postquantum models as resource theories~\cite{CoeckeSpekkens,SpekkensResource,GottesResource}. In popular accounts, the resource does not need to be seen as fundamental from the theoretical point of view and one does not need to establish an explicit relation between quantum technology and the underlying theory.

The method of popularization that starts from simple protocols is an efficient shortcut. It caters to the need to give a brief and comprehensible description of quantum technology. I am convinced, however, that the central challenge of popular and semi-popular accounts of quantum technologies is to not abandon the goal of explaining strange features of quantum theory, because the pragmatic approach possesses several important weaknesses.
The key one has to do with the fact that, while the lay audience is being told a pragmatic story about quantum technologies, it does not learn about what makes these technologies different from classical ones. Indeed, a pragmatic account of quantum technologies may not differ in structure from a pragmatic introduction to the new applications, e.g., of hydrodynamics or biotechnology. All of these technologies appear as yet another wave of cutting-edge developments that need to be brought into the public eye. In my view, this result cannot be deemed satisfactory. The public should not be left with a belief that quantum technologies are yet another feat in a long line of new technologies. The name ``quantum'' implies that something is dramatically different from the incremental development of many generations of classical technologies. It is imperative to explain this difference.
Thus an adequate popular account should aim at showing in a clear way, and perhaps explaining, the specific quantum leap in any new technology. A story that fails to convey a sense of what is profoundly different would fail
to do justice to the name: these technologies are \textit{quantum} and not merely \textit{new}. Their name has to be explained; but a pragmatic approach cannot do it.

Other disadvantages of the pragmatic approach can be grouped in three categories. Firstly, a hands-on method relies heavily on wonder and amazement to be produced in the audience. The  wonderful workings of a protocol or a device are presented as entirely new and seen never before. While this is often scientifically accurate, it does not help a layperson to place quantum technologies in the continuous history of ideas by connecting new information to previous knowledge. Secondly, this approach does not aim at understanding. A pragmatic heuristic only provides information about how one can use resources. Understanding the nature or the internal workings of these resources is not a goal, at least not a primary one. Thirdly, this approach cannot evoke in the audience a sense of participation or a first-person feeling of knowing, even if partially, how the quantum physicist works. Today, the society's judgment about new technologies is largely due to the methods and ways of virtue ethics~\cite{Swierstra2007,Davies2010,grinbaumGroves}. The figure of the scientist is at the center of such judgment. The risks and benefits of new technologies are still important but trust in science is not entirely due to an analytic calculation of the utility function. If the science-society interaction in the field of quantum technologies is to be successful, it cannot ignore this social reality. Hence the need to invent and to employ methods of popularization adapted to social demand. In a world where the public evaluates, not only the message it is given about the miraculous working of technology, but also the men and women who have made it, the research work of human scientists should be reflected upon as carefully and presented to the public as thoughtfully as one does when dealing solely with scientific or technological results.

\section{A narrative of entanglement}
\label{sect_entangle}

In 1935 Erwin Schr\"odinger called entanglement ``not one but rather \textit{the} characteristic trait of quantum mechanics, the one that enforces its entire departure from classical lines of thought''~\cite{schro35bis}. Thirty years later John Bell's work has put entanglement squarely in the center stage of quantum foundations. Surprising as it may sound, this central role of entanglement is still unrecognized in many popular texts or elementary textbooks on quantum theory. Superposition and the Schr\"odinger cat metaphor have proved effective and resilient to the attempts to replace them. But in the last twenty years Bell inequalities at last began to enter the mainstream of quantum physics (Figure~\ref{fig_Bell}). Gradually but inevitably, they are also making their way into the mainstream of educational and popular literature. At the same time, other notions are explored in the discipline of foundations, e.g., quantum contextuality~\cite{kochspeck65,KochSpeck}, showing that entanglement does not capture all the non-classicality of quantum theory. In quantum cryptography one witnesses a slow crystallization of new fundamental concepts like device-independence. In the last years the amount of work on such concepts has grown significantly (Figure~\ref{fig_stats}). It is not unfathomable that they will soon claim a place of their own in the popular and educational literature on quantum technologies.

\afterpage{%
    \clearpage
    \thispagestyle{empty}
\begin{landscape}
\begin{figure}\vskip -3.5cm
\includegraphics[scale=0.7]{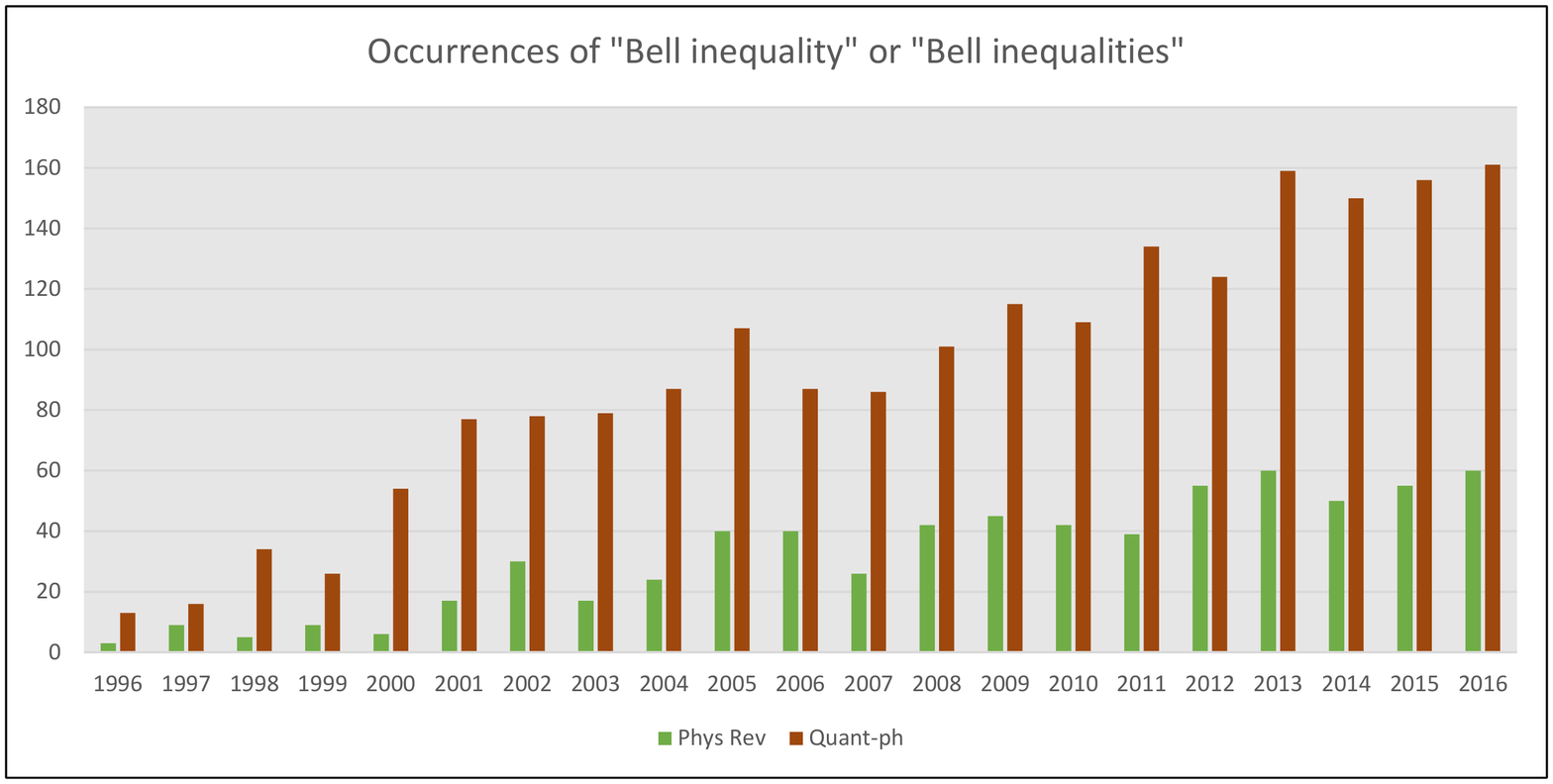}\vskip -1cm
\caption{Occurrences of `Bell inequalities' in \textit{Physical Review} articles and \texttt{quant-ph} preprints on \textit{arXiv}.}
\label{fig_Bell}\end{figure}
\end{landscape}
}
\afterpage{
\begin{figure}\centering
\includegraphics[scale=0.4]{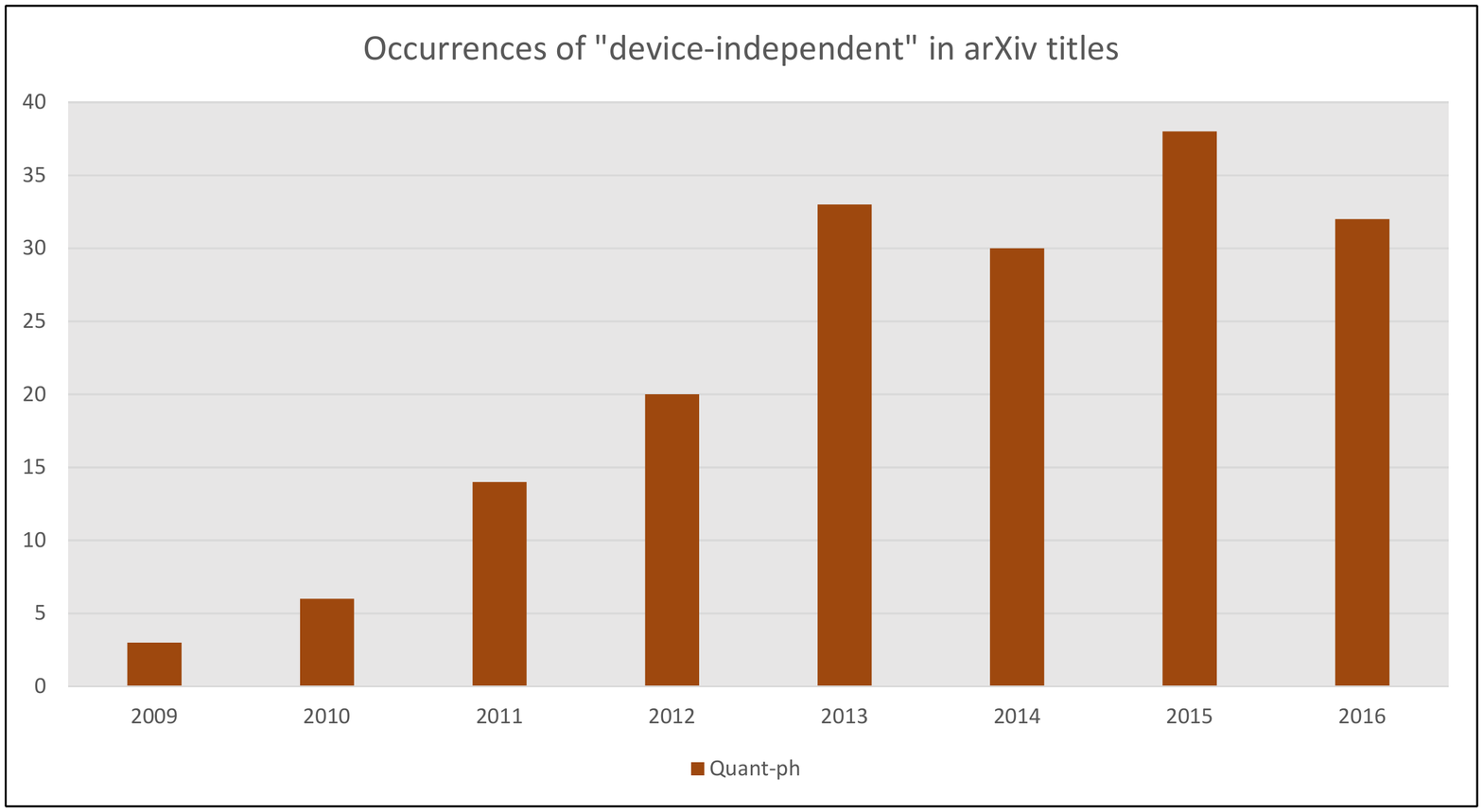}
\includegraphics[scale=0.4]{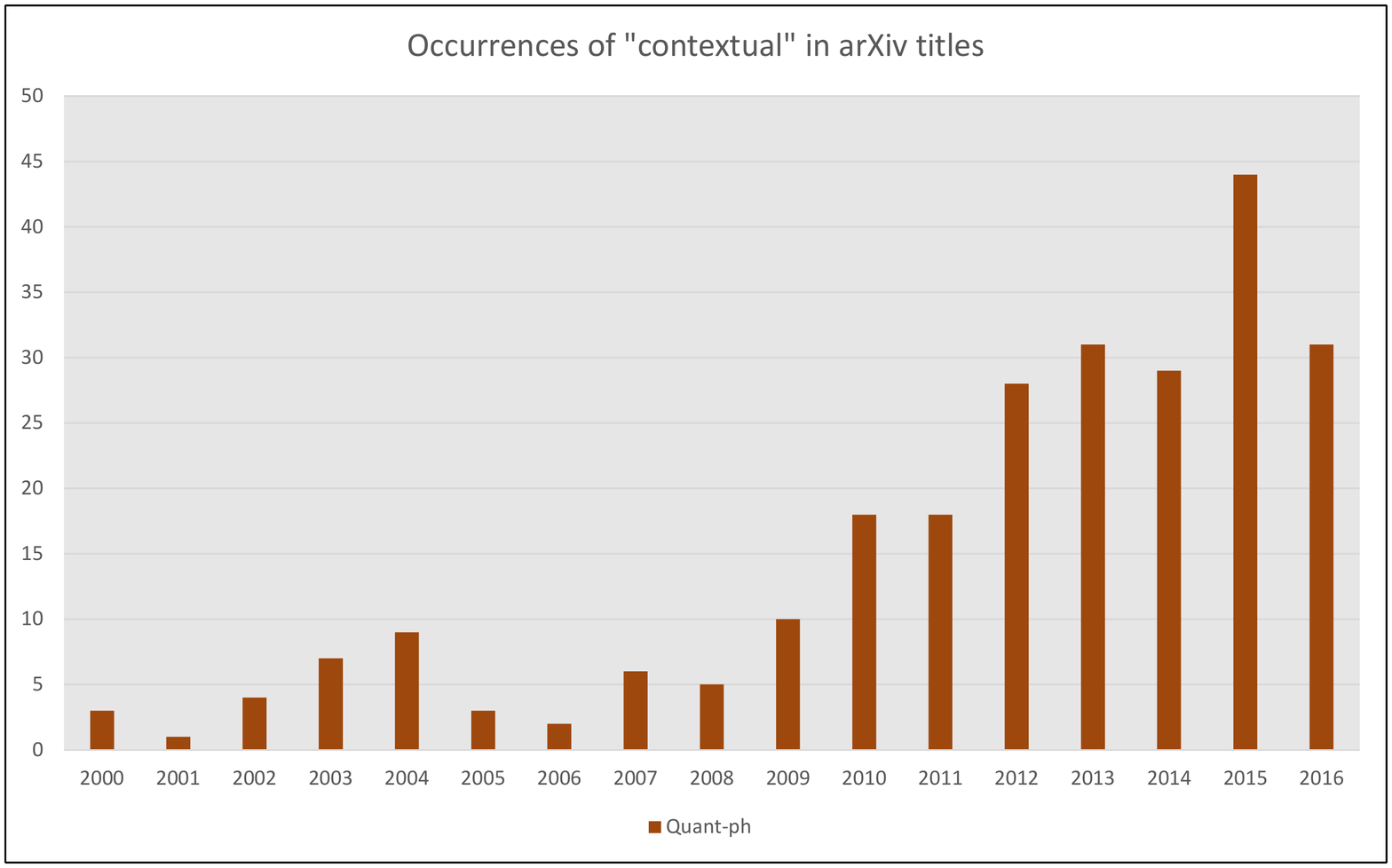}
\caption{Occurrences of ``device-inde\-pen\-dent''  and ``contextual'' in the titles of \texttt{quant-ph} preprints on \textit{arXiv}.}
\label{fig_stats}\end{figure}
}

Still, it is today beyond doubt that entanglement plays a crucial role in the explanation of the workings of a quantum computer or of a cryptographic device~\cite{RaussenBriegel,Eckert1991}. To introduce entanglement as effectively as the Schr\"odinger cat metaphor allows for the introduction of the superposition principle, it is important to find one or several metaphors capturing the essence of this concept. In what follows I will propose to position entanglement in a popular account as the last episode in a history of ideas about the composition of entities~\cite{grinbaumBook}. That entanglement is a tricky and unusual form of the composition of parts was clear already for Schr\"odinger: ``The best possible knowledge of a whole does not necessarily include the best knowledge of all its parts, even though they may be entirely separated and therefore virtually capable of being `best possibly known,' i.e., of possessing, each of them, a representative [state -- AG] of its own''~\cite{schro35bis}. By referring to the history of ideas, I submit that it is not enough to try to explain to the public the rational argument behind the workings of quantum technologies. No such explanation is truly possible in a popular account, which does not make use of the mathematical formalism of quantum theory. The pragmatic argument discussed in the previous section produces a narrative built on a short logical link between an available resource and the technological feat which it enables. This narrative provokes wonder and amazement more than it gives rational understanding. As this methodology becomes mainstream, it pushes one to reflect on other possible stories in the popularization of quantum theory.

The conundrum about impossible explanation has been known for a long time and is not limited to quantum theory. A typical solution employs metaphors to replace the incomprehensible mathematics. A metaphor is formulated in the ordinary language and appeals to laypersons far more directly than a mathematical formula. One such metaphor, the Schr\"odinger cat, is famously spectacular. The image of the Schr\"odinger cat counts among major visualizations of quantum mechanical entities~\cite{Mashh1,Mashh2}. It empowers an emotional narrative that introduces vector spaces and linear algebra. The outstanding results that this metaphor achieves are due, not so much to its faithfulness to quantum formalism, but to what lies beyond the scope of analogy, namely the human affectionate attachment to cats. The Schr\"odinger cat story is remarkable because it unites several functions: it conveys an approximate sense of what is going on physically, sends a metaphoric message about the strangeness of quantum theory, and achieves its goals thanks to emotional appeal. Whether all these traits can be found in a story about entanglement remains to be seen. It is, however, indubitable that narratives are critically important for public perception of science and for the relations between science and society~\cite{Dupuy2010,grinbaumGroves}.

To build a narrative about entanglement, I first propose to gently take the audience out of its comfort zone by discussing the logical awkwardness of common-language examples of composition:
\begin{description}
  \item[Glasses of orange juice] When two glasses of orange juice are put side by side, they remain two glasses; but when the juice is poured from the second one into the first, we are left with just one glass of orange juice. The second glass is simply a glass: it does not contain juice anymore. Now there is only one system called `glass of orange juice.' Thus the workings of language suggest that the composition of entities may be less straightforward than a matter of simple numeric counting.
  \item[Rotting bananas] We call a system `banana' as long as it recognizable as one by its shape and color. If, for example, a banana is left outdoors for a sufficiently long time, it rots and ceases to be recognizable. We do not call it a banana anymore. Thus dynamical temporal processes suggest that the identity of systems, exemplified in their names, may not be stable in time.
\end{description}
Such trivial examples bring home the idea that, above and beyond the common-language descriptions of the composition rules of systems, a theoretical analysis is necessary to put some order in this notion. As a second step in popularization, a brief history of composition may include the following two episodes~\cite{grinbaumBook}.

The initial episode has to do with the efforts in Greek philosophy to characterize different types of mixtures and compounds, as exemplified in the notions of \textit{synthesis} ($\sigma\acute{\upsilon}\nu\theta\varepsilon\sigma\iota\varsigma$), \textit{mixis} ($\mu\acute{\iota}\xi\iota\varsigma$), and \textit{krasis} ($\kappa\rho\acute{\alpha}\sigma\iota\varsigma$) in Aristotelian and Stoic physics~\cite{SamburskyStoics,GouldChrys}. To illustrate these notions, one may use a glass of wine and a vessel filled with water and mix them, then discuss various interesting properties of the mixture. Aristotle, Chrysippus and others analyzed actual or potential reversibility of the fusion, spatial location of its components, a question whether blending between water and wine is total or the two remain positioned side by side, and so forth.

The next episode in this brief history of composition comes from the theological debate in early Christianity about the type of union of divine entities. Between the third and the ninth centuries there existed a debate on how one should combine the two natures of Christ or the three Persons of Trinity. Many a theologian participated in this discussion before final orthodoxy was formulated and accepted following the teaching of John of Damascus. A particular type of composition chosen by the Christian doctrine, called \textit{perichoresis} ($\pi\epsilon\rho\iota\chi\acute{\omega}\rho\eta\sigma\iota\varsigma$) in Greek and \textit{circumincessio} in Latin~\cite{Cross,Durand}, owes much to the Stoic notion of \textit{krasis} and possesses several unintuitive, seemingly paradoxical properties. Surprisingly, perhaps, these are not dissimilar to the non-classical traits of quantum entanglement~\cite{grinbaumBook,Wegter}. For example, it would be a theological heresy to stipulate that the composition of the two natures of Christ has appeared at some point in time. According to doctrinal teaching, this union does not have a temporal beginning or an end, for the ``system'' is ``defined'' atemporally and exists outside and beyond worldly dynamics. Temporality, therefore, is not a property of christological perichoresis; it is also not a trait of quantum entanglement. The latter is defined in the absence of dynamical evolution as a purely algebraic property in the Hilbert space. To illustrate this point to an audience that cannot comprehend the mathematical notion of Hilbert space, one may resort to analogy with theology: perichoresis possesses a similar trait in a rigorous albeit not a mathematical theoretical framework. To continue the example, it would be a theological heresy to surmise that the union of the three Persons of Trinity has spatial bounds or is located somewhere. Instead, according to doctrine, it permeates space entirely and is not circumscribed. One may use this trait to illustrate Schr\"odinger's point that entanglement in composite quantum systems is not a three-dimensional notion: even if the subsystems are ``entirely separated,'' they are described by a single wavefunction and form a single quantum system. Thus the non-spatial character of trinitarian perichoresis provides a useful non-mathematical way of introducing this counterintuitive feature. It renders less unfamiliar the property of entanglement that has no boundary or location in Euclidian space.

Parallels like the one between entanglement and perichoresis serve two purposes. Firstly, in their metaphoric capacity they enable a non-mathematical introduction of the concept of entanglement. Secondly, they render the strange properties of entanglement less unique and slightly less unusual. When new knowledge is presented to the lay public in the narrative form, it is critical that they be able to connect it with their previous knowledge. A narrative must be contextualized, or else it will not be `domesticated' or brought into broader culture~\cite{Wynne}. By linking the mathematical properties of entanglement with other episodes in the history of ideas, one gives quantum theory and quantum technologies a place in history and a place in culture. It is hard to underestimate the importance of doing so.

Quantum mechanics is typically accompanied by an aura of absolute novelty and total strangeness. For a popular account, it is crucial to dispel this aura. Ultimately, a layperson must be reassured: even if quantum technologies are indeed strange, even paradoxical, they are less otherworldly than one may have feared. They are not divine or demonic, for they have parallels in human culture and in the history of ideas. Bringing them into everyday life would then become more routine and acceptable.

The analogy with perichoresis is perhaps less emotionally attractive that the parable of the Schr\"odinger cat but it is equally human-centered. The emotional appeal of cats is replaced with a widely known episode of human intellectual history. The theological simile achieves a similar goal in the popularization of entanglement as the cat narrative did for superposition. This brings us forward from an early stage of popularization of quantum mechanics to new opportunities offered by the need to communicate about quantum technologies.

\section{Beauty and trust}
\label{sect_myth}
Narratives enable yet another crucially important achievement in the popular accounts of quantum technologies: they can be employed to transmit a sense of beauty of physical theory. One should not belittle the place of beauty in the scientist's thinking, emphasized by Russell:
\begin{quote}
Mathematics, rightly viewed, possesses not only truth, but supreme beauty, a beauty cold and austere, like that of sculpture, without appeal to any part of our weaker nature, without the gorgeous trappings of painting or music, yet sublimely pure, and capable of a stern perfection such as only the greatest art can show.~\cite[p.~49]{RussellMath}\end{quote} Like other physicists, quantum theorists develop an intuitive aesthetic heuristic by working with the mathematical formalism of quantum theory. Mathematical reasoning leads to the emergence of a feeling of formal elegance, which subsequently serves as a thinking aid.

If aesthetic intuition is employed as a guide for making discoveries in physics, it can be extraordinarily fruitful but also misleading~\cite{grinbfine}. For the purposes of my argument, suffice it to say that beauty and truth, as well as beauty and good, are distinct categories, in particular in their application to physics and to technology. There is no logical link between these notions: the beautiful may be false or evil, and the true may be ugly. However, beauty is essential to the physicist's work in quantum theory. In the 1930s Einstein expressed his conviction that ``pure mathematical construction enables us to discover the concepts and laws connecting them, which give us the key to the understanding of the phenomena of Nature''~\cite{e3}. A younger Einstein was more critical of the value of formal arguments as a guide to physical truth but still emphasized their importance for building the theory: formal arguments ``may be valuable when an \textit{already found} [his emphasis -- AG] truth needs to be formulated in a final form, but fail almost always as heuristic aids''~\cite{e2}. Even those authors who, like Abraham Pais, side with the younger Einstein's more critical position, admit that ``it is true that the theoretical physicist who has no sense of mathematical elegance, beauty, and simplicity is lost in some essential way''~\cite[p.~172]{Pais}. I believe that, in order for the public not to be lost about quantum theory in some essential way, it is imperative that they experience beauty. How this can be achieved given that mathematical formalism cannot be used in popular accounts is the \textit{hard problem} of the interaction between science and society on the subject of quantum technologies.

Beauty is not limited to the mathematical formalism of quantum theory. It also characterizes the ingenious experimental setups needed to test quantum phenomena. But small scales and large scales have been explored for decades. The public is familiar with the story about `two infinities' as two feats of physical experimentation. The familiarity of the story makes it less attractive to the audience. Additionally, in quantum technologies, it is particularly hard to draw from this source of beauty due to the non-bespoke nature of many experiments. Quantum cryptographic devices are often black boxes, the internal workings of which remain a commercial secret. In quantum computing, all currently available hardware that exhibits computational advantage over classical computers is also proprietary. Quantum optics, too, uses tabletop devices with relatively mundane technology of mirrors and lasers. All of this does not let quantum technologies to easily offer themselves as a source of wonder. Black boxes cannot be opened before the public eye, yet something spectacular needs to be shown to the lay audience to replace secrecy by another impression. The non-breathtaking experimental setups that realize novel physical phenomena remain inherently beautiful for the experimenter; conveying this sense of beauty is an important challenge on a par with the difficulty of transmitting mathematical beauty of the theory.

It is unavoidable to resort to metaphors in popular accounts of quantum theory. Whichever common-language metaphor one uses, it necessarily leads to the loss of the sense of mathematical elegance that the physicist experiences when working mathematically. If beauty cannot be so conveyed by a metaphor, however, it can still be embedded in the narrative that encompasses that metaphor. This does not occur with narratives that are rational stories, like the Schr\"odinger cat parable. But if a narrative can draw from its own source of beauty, inherent to the literary genre of that narrative, then it may provoke a feeling of elegance. This sense of beauty is not based on mathematics but it is still a fully-fledged aesthetic notion. Russell compared mathematical beauty with the beauty of poetry: ``The true spirit of delight, the exaltation, the sense of being more than Man, which is the touchstone of the highest excellence, is to be found in mathematics as surely as in poetry. What is best in mathematics deserves not merely to be learnt as a task, but to be assimilated as a part of daily thought\ldots''~\cite[p.~49]{RussellMath}. I propose to choose myth as another type of narrative capable of provoking an aesthetic feeling and transmitting the idea that beauty is part of the scientist's daily thought.

The sample narrative about entanglement proposed in Section~\ref{sect_entangle} is built around several episodes from the history of religion and theology. These are not stories containing testable predictions subject to experimental science; they originate in myth and, though rigorous in their own way, they differ from realist or hyperrealist parables of the Schr\"odinger cat type. This profound difference is often disturbing for the public (laypersons tend to believe that modern science has nothing to do with myth or theology) but it also opens new opportunities for popular science authors. Myths are built following their own logic and express a different kind of beauty via common language. Lay persons experience this beauty when they are confronted with a mythological story. It is not unfathomable that such a narrative, when it is carefully crafted by popular science authors, may both be scientifically informative and retain its original mythological beauty.

\section{Conclusion}
Introductions to quantum technologies for the lay public usually follow a pragmatic approach rooted in the information-theoretic reconstructions of quantum theory. Their logical form and simplicity are the two aspects that make them attractive. Technology is presented in one or a few steps that take the audience from a simple protocol or a resource to an application that performs useful work or computation. Many other aspects of the workings of quantum technologies are often hidden in a `black box': a theoretical construct needed to single out one particular feature that will be put in the center of the story, or an actual device whose internal structure cannot be publicized due to secrecy. In both cases the blackbox approach proves to be an effective hands-on method. Yet it also has several weaknesses that may prevent the public from building trust in quantum technologies. To go beyond pragmatism and build trust in quantum technologies, I suggest that it is imperative to appeal to the figure of the quantum physicist and, in particular, to his or her professional aesthetics. Constructing a narrative that conveys scientific content as well as provoking a feeling of beauty is the hard problem in the relations between science and society. In the case of entanglement, mythological stories about the composition of parts provide a useful set of analogies on the way to solving this problem. They are complementary to the pragmatic approach to popularization and, crucially, they enable an explanation of the specific strange properties of quantum technologies. Pragmatism and beauty, in conjunction, show a promising way forward in the development of the science-society interaction in the field of quantum technologies.

\singlespacing\footnotesize
\bibliographystyle{habbrv}

\end{document}